\documentclass{article}
\usepackage{graphicx}
\usepackage[below]{placeins}
\usepackage{psfig}
\usepackage{epsf}

\newcommand{\beq}{\begin{equation}}
\newcommand{\eeq}{\end{equation}}
\newcommand{\beqa}{\begin{eqnarray}}
\newcommand{\eeqa}{\end{eqnarray}}

\begin{document}

\begin{titlepage}

\begin{flushright}
LU TP 01-09\\
March 15, 2001
\end{flushright}

\vspace{.25in}

\LARGE

\begin{center}
{\bf Topological Properties of \\ Citation and Metabolic Networks}\\
\vspace{.3in}
\large

Sven Bilke and   
Carsten Peterson\footnote{\{sven,carsten\}@thep.lu.se}\\ 
\vspace{0.10in}
Complex Systems Division, Department of Theoretical Physics\\ 
University of Lund,  S\"{o}lvegatan 14A,  S-223 62 Lund, Sweden \\
{\tt http://www.thep.lu.se/complex/}

\end{center}
\vspace{0.25in}

\large
{\bf Abstract:}

Topological properties of "scale-free" networks are investigated 
by determining their spectral dimensions $d_S$, which reflect a diffusion 
process in the corresponding graphs. Data bases for citation 
networks and metabolic networks together with simulation results from 
the growing network model \cite{barab} are probed. For completeness and 
comparisons lattice, random, 
small-world models are also investigated. We find that $d_S$ 
is around $3$ for citation and metabolic networks, which is significantly 
different from the growing network model, for which $d_S$ is approximately 
$7.5$. This signals a substantial difference in network topology despite 
the observed similarities in vertex order distributions. In addition, the 
diffusion analysis indicates that whereas the citation networks are tree-like 
in structure, the metabolic networks contain many loops.

\begin{center}
Submitted to {\it Physical Review {\it E}}
\end{center}

\large

\vspace{0.8in}

{\it PACS numbers}: 05.10.-a, 05.40.-a, 05.50.+q, 87.18.Su

\end{titlepage}

\large

\newpage

\section{Introduction}
There has recently been an upsurge of interest in so-called 
scale-free networks, where the vertex order, or degree of connectivity per 
node $k$, follows a power-law distribution
\beq
P(k) \sim k^{-\gamma}
\label{s-f}
\eeq
for $k> \langle k \rangle$. This is in contrast to the exponential 
suppression expected from randomly wired networks. For a number of 
real-world networks like social networks, powergrids, citation networks, 
the World Wide Web and metabolic networks \cite{barab_wit}, the scale-free 
behaviour of Eq. (\ref{s-f}) has been observed with exponents $\gamma$ in 
the range 2-4 \cite{barab}. 
The nature of all examples above is that they originate from a 
growing process. Hence, it appears natural to develop a model of 
growing network, which was indeed done in \cite{barab}. 

The degree of connectivity is a measure of local properties of the 
networks. Less attention has been payed to more global properties 
for the above examples except for general features like network diameters. 
Global measures could be important discriminants of network properties, 
in particular when it comes to narrow in on potential underlying models.  

In this paper, we analyze topological properties of networks, both of 
synthetic and real-world nature, by extracting  spectral dimensions 
$d_S$ using a random walk procedure, from which the return-to-origin probability 
is estimated. The focus is on scale-fee networks; citation networks, 
metabolic networks and the growing network model \cite{barab} are 
investigated. For completeness and comparisons, we also extract the 
$d_S$ from simple three-dimensional regular lattice networks, 
random networks and small-world networks~\cite{sw}.

We find that the citation and metabolic networks are quite low-dimensional 
with $d_S$ around $3$, whereas for the growing network model $d_S$ is  
approximately $7.5$. The latter turns out to be more in parity with 
what characterizes random and small-world networks. As a consistency check,  
the dimension of regular lattice networks is also determined, which 
comes out as expected. One concludes that the spectral dimension offers a 
powerful and additional measure to $\gamma$ in Eq. (\ref{s-f}) when it 
comes to characterizing network topologies. Furthermore, the diffusion process 
underlying the extraction of $d_S$ hints upon differences in the 
citation and metabolic network topologies; the former is tree-like 
whereas the latter is rich in loop structures.

This paper is organized as follows: In Sect. 2 we describe the method 
for extracting the spectral dimension and Sect. 3, 4 and 5 contain our 
investigations of the synthetic, citation and metabolic networks   
respectively. A summary can be found in Sect. 6.

\vfill\eject

\section{Spectral Dimension}
Our method to probe the topological properties of an interaction 
network, which is of more global nature than the degree of connectivity, 
is based upon the diffusion of a test particle in the metric space defined 
by the graph. In a continuous space with a fixed smooth metric,
the diffusion equation has the form
\beq
\frac{\partial}{\partial t} K_g (\epsilon, \epsilon _0, t) =
 \Delta _g  K_g (\epsilon, \epsilon _0, t).
\eeq
Here $t$ is the diffusion time, $\Delta _g$ is the Laplace operator 
in the metric $g$ and $K_g (\epsilon, \epsilon _0, t)$ is the 
probability density to diffuse from $\epsilon _0$ to $\epsilon $ 
in time $t$. For small $t$ it is well known~\cite{return}, that the average 
{\em return probability} has the following asymptotic expansion
\beq
\label{dc}
K_g(0,0,t)  \sim t^{-d_S/2}  
\eeq
with 
\beq
\label{asynp}
 K_g(\epsilon,0,0) =  \delta (\epsilon)  
\eeq
and where the power $t^{-d_S / 2}$ reflects the dimension of the network. 
        
In the spirit of these equations we extract the {\em spectral
dimension} of the geometry  defined by the interaction networks 
considered in this work. To this end,  we use the transition matrix
%
%
\beq
C_{ij}   = \frac{J_{ij}}{k_i}   
\eeq
with
\beq
  J_{ij} = \left \{ 
             \begin{array}{ll}
		 1 & \mbox{if $i$ and $j$ are neighbours} \\
		 0 & \mbox{otherwise}
	     \end{array}
	   \right .
\eeq
as the discrete version of the Laplace operator $\Delta _g$. The number $k_i$,
counts the number of links connected to vertex $i$ -- the vertex-order. In a 
simulation of the diffusion process defined by the transition matrix, the 
probability to return, $P$, after $t$ steps to the origin $\delta _i$ is 
then measured. More precisely, we choose a random subset 
$\{ \delta _i \}, i= 1 \cdots N$ of vertices and extract 
\beq
<P(t)> = \frac{1}{N} 
          \sum _i ^N \left < \delta _i | C_{i,j}^T | \delta _i \right >.
\label{discRP}
\eeq
We then fit the resulting distribution to the asymptotic form
in eq. (\ref{dc}). For large $t$, Eq.~(\ref{dc}) is dominated by the 
eigenvector of $C_{ij}$ with eigenvalue $\lambda = 1$; the diffused particles  
reach an equilibrium distribution and Eq. (\ref{dc}) does not hold. 
Also, for too small $t$ the assumption of 
a smooth metric is not justified. Our extraction procedure is therefore 
somewhat more elaborate to account for these effects. A sliding window method 
is used, where the window is chosen such as to minimize the 
standard deviation per point used in the fit. In other words,  we fit to the 
part of the distribution, which is closest to the functional form assumed 
in this procedure. Depending upon $d_S$, window sizes might vary from a few 
time points to the entire range.

 \section{Synthetic Networks}

Next we extract $d_S$ for a few different types of synthetic networks with 
sizes up to around 30000 nodes in order to compare with typical sizes of 
citation networks. In addition, for all networks the average geodesic distance 
(diameter) $R$ is computed.

\begin{figure}[htb]
\vspace{0.3in}
\centerline{\psfig{file=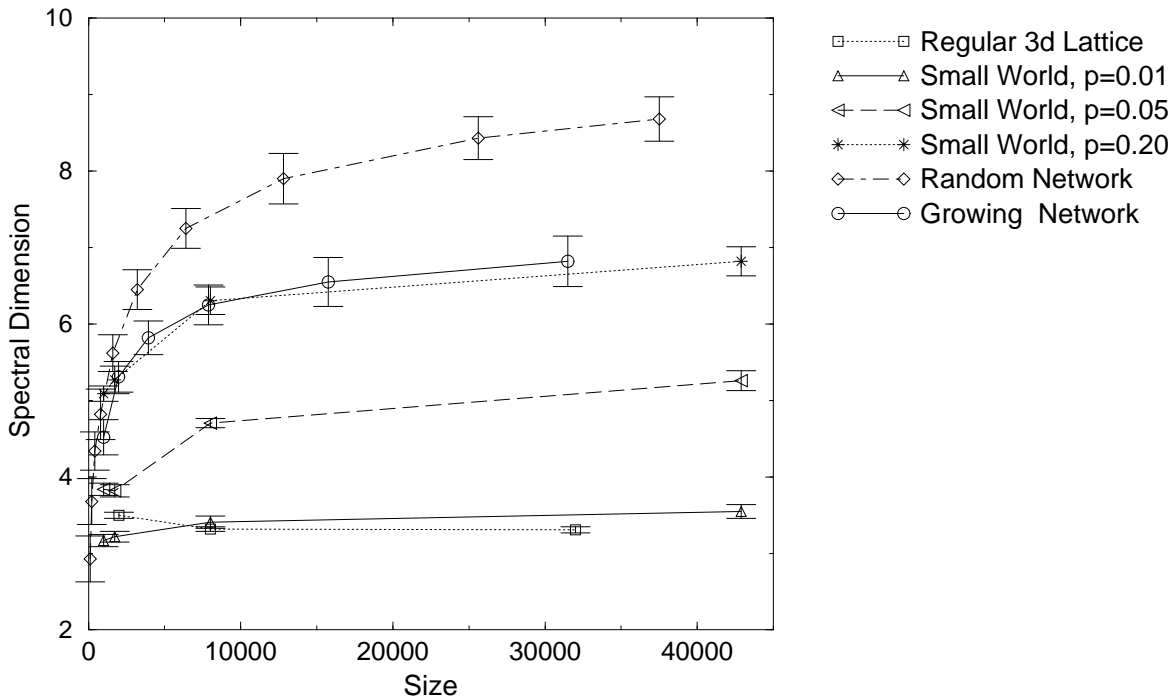,angle=270,height=9.5cm}}  
\caption{Spectral dimensions $d_S$ as functions of size for synthetic networks.}
\label{DS}
\end{figure}

{\bf Lattice Networks.} We use a simple {\em square} lattices with
dimension $d=3$. For these it is only possible to return to the 
origin after an even number of steps. Therefore $P(t) = 0$ for odd $t$. 
In terms of the transfer matrix eigenvectors this means that the symmetry
of the square lattice leads to an eigenvector with $\lambda $ exactly
$-1$. The corresponding eigenstate does not decay for large $t$ and yields 
destructive interference with the $\lambda = 1$ eigenstate 
after an odd number of steps. When extracting the spectral dimension
$d_S$ from the data we therefore omit the odd time-points. 
As can be seen from Fig. \ref{DS}, $d_S$ is slightly above 
the expected value $3$, the dimension of the lattice. We have repeated
this experiment for a regular lattice including diagonals, hence allowing 
for closed loops with an odd number of steps. Although local properties
of the graph are considerably changed since the number of links emerging 
from each vertex is increased from $2d$ to $2d + d(d-1)$, we still 
extract the same value for $d_S$ within errors. This is a further justification
of omitting the data at odd time points for the square lattice and
emphasizes that the spectral dimension is sensitive to global 
geometric aspects rather than to local details.    

{\bf Small-world Networks.} We generate small-world networks \cite{sw} from
the lattice network  described above by rewiring the edges in
the regular lattice to a randomly chosen vertex with probability 
$p$= $0.01$, $0.05$ and $0.2$  respectively. This generates three sets 
of models for each system size. Not surprisingly, these networks end up
with spectral dimensions (see~Fig.~\ref{DS}) in between the regular lattice and the 
homogeneous random networks described next.

{\bf Random Networks.} Homogeneous random networks are generated as  
directed networks, where each vertex has two inputs from other vertices 
choosen at random. In the analysis the graph is considered undirected 
by ignoring the orientation on the edges connecting the vertices. 
Furthermore, the analysis is focused on the largest cluster; the largest 
connected part of the resulting graph. The spectral dimension 
$d_S \approx 8.5$ (see~Fig.~\ref{DS}) is relatively large. This agrees 
qualitatively with the observation that a characteristic length-scale, 
e.g. average distance between vertices along links, grows slowly with 
the volume. It also signals a large dimensionality for this type of graph. 
It is interesting to note that again a distinct even-odd disparity emerges; the 
amplitude after an odd number of steps is almost zero. This indicates that 
the random network is dominated by a tree-like structure, the number of 
loops (at least of odd length) is negligibly small compared to the amplitude 
obtained from backtracking the same path.

\vfill\eject

{\bf Growing Network Model.}
In this model \cite{barab} one at each time step $t$ adds a node, which 
connects to $m$ existing nodes ($i$) with a probability $\Pi_i$ given by 
\beq
\Pi_i = \frac{k_i^{\beta}}{\sum_jk_j^{\beta}}
\label{grow}
\eeq
where $k_i$ is the connectivity of node $i$ and $\beta$ is a 
parameter. In \cite{barab} the $\beta$=1 case was 
investigated both analytically in the $t \rightarrow \infty$ case 
and numerically for finite~$t$. With the approximations involved 
one arrived at $\gamma=3$ in the analytic approach and 
$\gamma=2.9 \pm 0.1$ from simulations. For these networks,  
$R$ is approximately  $6$ for large system sizes. 
%

\section{Citation Network}
%
\begin{figure}
\centerline{\psfig{file=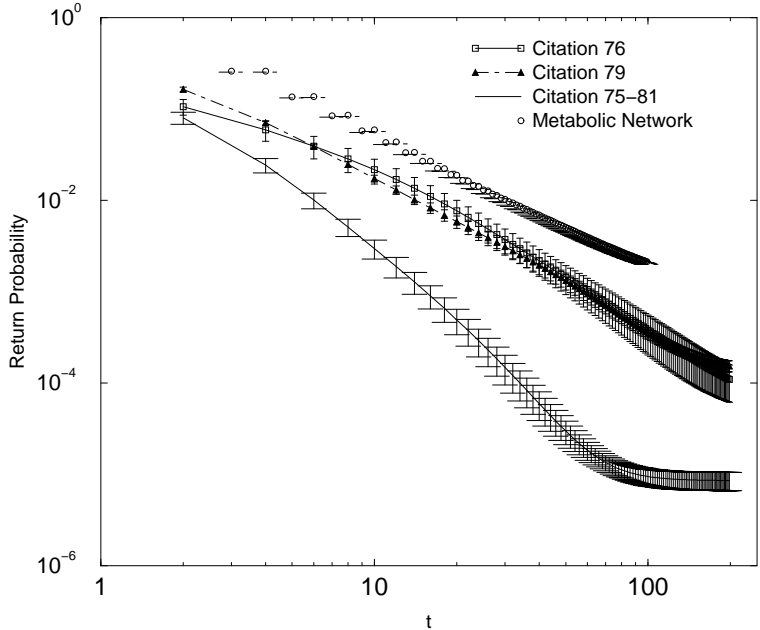,angle=270,height=8.5cm}}  
\caption{The return probability ($P$) as a function of time steps $t$ for
citation networks with citing  publications from a single years, 1976 and 1979
respectively, and for  a six year period (1976-81). Same quantity for the
metabolic network.  To maintain readability, $P$ for the metabolic network is
multiplied with a factor $5$.} 
\label{f:RP}
\end{figure}
In a citation network the links and nodes correspond to the citations and the
publications (citing and cited) respectively. We use the SPIRES data base
\cite{spires} for our citation network studies. This data base is limited to
high energy physics publications but is not confined to articles that have
been published in refereed journals -- citations to and from conference
reports are also present. From the data we construct citation networks, which
we treat as undirected, not fully connected graphs. By considering graphs
generated by publications in a certain year or  time span and the cited papers
(regardless of publication year) we obtain a whole set of graphs. In these not
fully connected graphs we focus on the largest connected cluster, the
corresponding  sizes are shown in Table \ref{t:spires}. When computing the
connectivity distribution of the nodes, we confirm the power law distribution
(Cf.~Eq.~(\ref{s-f})) already observed in citation networks \cite{redner} with
$\gamma=2.7$.


In Fig. \ref{f:RP} we plot the probability to return to the origin 
after $t$ time steps for a few typical networks. It is interesting to note
that, within errors, the return probability for the 76 and 79 networks
are in very good agreement indicating universal topological
properties.  This holds true also when comparing all of the
other one-year networks with the exception of the $75$ and the $87$ networks 
(not shown), which cannot be fitted to Eq. (\ref{dc}). In these years the 
connectivity does probably not generate a homogeneous metric, because the 
geometry is dominated by two or more large clusters, which are interconnected 
by only a few links.  

For the networks composed by several years we again observe a remarkable
universality when comparing different time-periods. However, the 
distribution observed
is considerably different from the one-year distributions. The reason 
presumably is a maturation of the citation networks leading to a tighter
interconnection of the central cluster. In turn we observe  larger dimension
as can be seen from the steeper slope of the corresponding graph in 
Fig. \ref{f:RP}.
These issues for the citation networks have been subject to further 
investigations \cite{bilke}.

\begin{table}[htb]
\begin{center}
\begin{tabular}{|cc|cc|}
\hline
Year& Largest Cluster &  Year  & Largest Cluster \\
\hline 
1975	& 20931 & 1983	& 32752 \\
1976 	& 22969 & 1984	& 34558 \\ 
1977	& 23936 & 1985	& 37020 \\
1978	& 26038 & 1986	& 39962 \\
1979	& 27055 & 1987	& 44392 \\
1980	& 28045 & 1988  &  47290 \\
1981 	& 29309 & 1989  &  45549 \\
1982	& 31516 & 75-81 & 98104  \\
\hline
\end{tabular}
\caption{Largest cluster sizes in the SPIRES data base for the years 
1975-1989.}
\label{t:spires}
\end{center}
\end{table}

\section{Metabolic Network}
Another real-world interaction network is the metabolic network found in 
living cells. In these networks, substrates are treated as vertices, 
while chemical reactions connecting substrates and educts are treated as 
directed links. Recently it has been demonstrated that for this type of 
networks the connectivity distribution obeys a power-law behavior 
(Cf. Eq.~(\ref{s-f})); these networks are scale-free with respect to the 
order distribution \cite{barab_wit}.  

Using data from the EMP-Project \cite{EMP} we constructed a network including
all reactions found in the database without taking into account species and
cell locations. As above, we neglect the  orientation of the resulting graph in
the analysis. The spectral dimension observed for this network is $d_S \approx 
2.8$. This is a surprisingly small dimension taking into account the 
observation \cite{barab_wit} that the average distance between  vertices on
this graph does not grow with the graph-size, implying a very large, possibly
infinite dimension.  Also, comparing $d_S$ with the corresponding value for
the growing model scale-free network analyzed above, the difference is
remarkable.  This indicates that, 
although the metabolic network and the scale-free network are similar on a local
(vertex order) scale, the more global topological properties are very 
different. Another important difference arises from the observation that the
return probability for odd path-lengths do not vanish. This means that the
metabolic networks consist of a very large number of closed loops. The graph
has, in contrast to  the random, growing model and citation networks, {\em
not} a dominantly tree-like structure. We interprete this as an indication of
a built-in stability of metabolic networks with respect to small modifications.

\begin{figure}
\centerline{\psfig{file=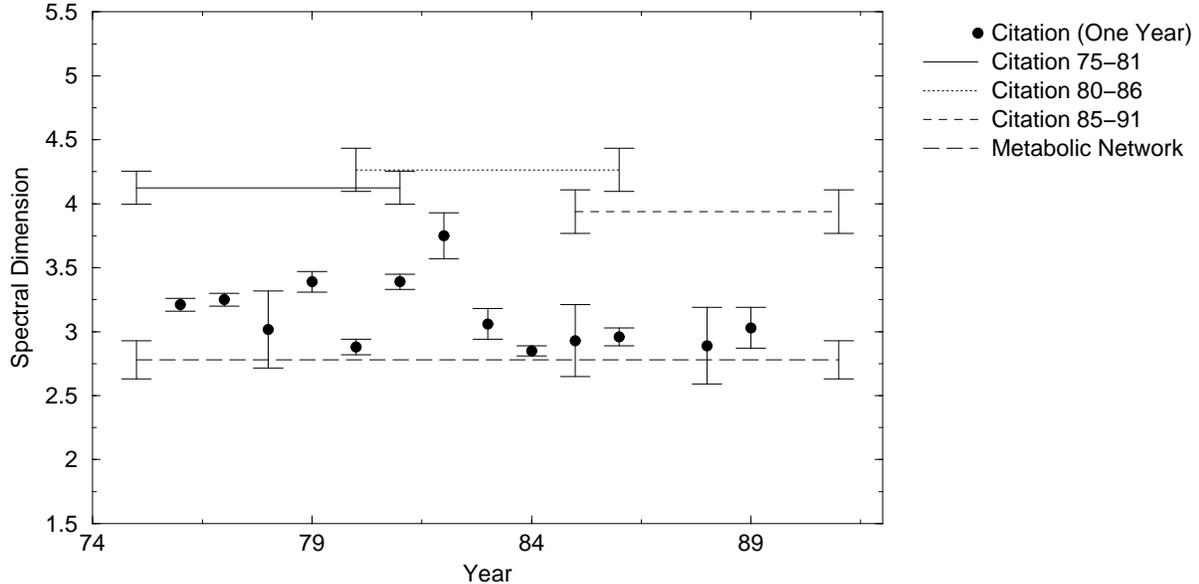,angle=270,height=8cm}}  
\caption{The spectral dimension $d_S$ for the citation networks for papers
         citing in one year and over six-years periods. For comparison, 
         $d_S$ for the metabolic network is also shown. } 
\label{f:DSR}
\end{figure}

\section{Summary and Outlook}

In table \ref{t:results} the results for $d_S$ are summarized for the 
different networks. Also shown here are the average (geodesic) distances $R$. 
For the regular lattice, the spectral dimension reproduces 
quite well the dimension of the underlying lattice. Depending upon the 
probability to rewire a connection ($p$), the small-world networks takes on 
values between the regular lattice ($p=0$) and the random network ($p=1$), 
which has the largest spectral dimension of all networks probed here. 
One might have guessed that the growing network has an even larger dimension
-- the average distance for a network with the same size is smaller.
However, in contrast to  the geodesic distance, which only counts the shortest paths, 
the spectral dimension takes into account all possible paths of a given 
length. This means that only a few links, in extreme cases even a single link, 
can considerably change the geodesic distance, while the spectral dimension 
probes larger parts of the geometry and is therefore only slightly affected. 
For example, consider  the behavior of the average distance for 
small-world networks with different probabilities $p$ to rewire the underlying 
three dimensional regular structure (table {\ref{t:results}). Even for 
$p=0.01$ with the resulting geometry still essentially a regular three 
dimensional lattice, the geodesic distance jumps to less than one half of 
the value for the regular case, while the spectral dimension is changed only 
moderately.  

\begin{table}[htb]
\begin{center}
\begin{tabular}{|l|cccl|}
\hline
Network                 & $N$ &   $R$   &  $d_S$  & $P_{odd}/P_{even}$ \\
\hline 
Regular Lattice (d=3)    & 32000 & 14.3 & 3.1  & 0 (1 with diagonal) \\
Random Network          & 32000 &   8.2  & $\approx  8.5$ & 0  \\
small-world Network ($p=0.01$)  & 32000 &   $5.9 $ & $ 3.6 $ & 0\\
small-world Network ($p=0.05$)  & 32000 &   $5.7$ & $5.3 $ & 0\\
small-world Network ($p=0.2$)   & 32000 &   $5.3$ & $6.8 $ & 0\\
Network Growing Model   & 32000 &   5.9  & $\approx  7 $   & 0 \\
Citation Network        & 20000 - 200000 & 6.3 - 5.6 &  $2.8 - 4.2$ & 0 \\
Metabolic Network       & 3800  &  3.1   & 2.8        & 1  \\
\hline
\end{tabular}
\caption{Largest cluster sizes ($N$), average distances [diameters] ($R$) 
and spectral dimensions ($d_S$) for different network models and real-world 
networks. Also indicated are whether the return probabilities ($P$) are 
substantial for even or odd steps.}
\label{t:results}
\end{center}
\end{table}

The spectral dimensions for the real-world networks, the citation 
and metabolic networks respectively, are strikingly similar in view of 
the much larger values observed for the network growing model and the 
random network. We interprete this as a sign of universality in the 
geometry of the real-world networks. The difference in $d_S$ when comparing 
to that of the network growing model is remarkable. As mentioned in the 
introduction, the latter reproduces quite nicely the power-law scale-free 
vertex order distribution observed for the real-world networks \cite{barab}. 
However, the difference in $d_S$ observed here indicates that the geometries 
of real-world networks are not fully described by the network growing model.  

Given the similarity in $d_S$ for the citation and metabolic networks, 
one might ask how similar they really are? An important difference can be 
observed by investigating the probability $P_{odd}$ to return to 
the origin after an odd number of steps. For the citation network, as 
for all the synthetic networks, one obtains $P_{odd} \approx 0$. This is in 
contrast to the metabolic network for which $P_{odd} \approx P_{even}$.
The case $P_{odd} \approx 0$ indicates a tree like geometrical structure 
with only few loops, where the return probability is dominated by the 
inverse of the forward paths. The regular lattice without diagonals seems 
to be a counter example.
The return probability $P_{odd}$ is exactly zero, but the geometry
is certainly not tree-like and there  exist many non-backtracking loops.
However, this lattice is a special case since its symmetry only 
allows for even step non-backtracking loops. It is extremely unlikely that
such an effect plays a role in any of the other lattices. 

Our results can be summarized as follows:

\begin{itemize}

\item The spectral dimension $d_S$, which reflects a diffusion process on networks, 
offers quite some promise in categorizing networks beyond what emerges 
from studying vertex order distributions.

\item In particular, we find that $d_S$ are very similar for citation 
and metabolic networks. This may indicate a universal behaviour of 
real-world networks. 

\item With respect to $d_S$, the growing network model is 
significantly different from citation and metabolic networks.

\item As a by-product when extracting $d_S$,  we find that whereas 
citation networks have tree-like structures the metabolic networks 
appear to contain many loops.

\end{itemize}

In addition, lattice, random and small-world networks were probed 
using $d_S$ with results exhibiting internal consistency of the 
method. 

\vspace{0.3in}

When finishing the write-up of this work, a paper with somewhat similar 
scope and philosophy but with employing eigenvalue analysis techniques 
instead was released; 
I.J. Farkas et. al., "Spectra of "Real-World" Graphs: Beyond the 
Semi-Circle Law" [cond-mat/0102335].

{\bf Acknowledgements:}
 
We are very much indebted to Louise Addis in the SPIRES-HEP data base group 
at the Stanford Linear Accelerator Center library for providing us with the 
SPIRES data base in a form that substantially facilitated this project. 
Alex Selkov from the EMP-Project supported us in acquiring the data 
for the metabolic network.
We have also benefitted from discussions with Bo S\"oderberg regarding 
eigenvalue properties of network diffusion.
This work was in part supported by the Swedish Foundation for Strategic 
Research and the Knut and Alice Wallenberg Foundation through the SWEGENE 
consortium.


\end{document}